\documentstyle[eqsecnum,aps,twocolumn]{revtex}

\def\mathbb#1{{#1}}
\def\Ll{\mathcal{L}}

\def\Nn{{\mathcal{N}}}
\def\Pp{{\mathcal{P}}}

\def\R{\mathbb{R}}

\def\d{d}
\def\dd{\,d}
\def\dx{\d x}
\def\dv{\d v}
\def\ddx{\dd x}
\def\ddv{\dd v}

\def\EXP#1{e^{#1}}

\def\DIFF{{D}}
\def\GRAD{\nabla}
\def\LAP{\Delta}
\def\DIV{{\mathrm{div}}}

\def\DIFF{\partial}

\def\COMMA{\,,}             % use for commas at the ends of formulae
\def\PERIOD{\,.}            % use for periods at the ends of formulae
\def\SEP{{\,|\,}}           % use for seperator in set defs, like {x\in R \SEP

\def\VIZ#1{(\ref{#1})}      % use for references to formulae
\def\LATT{{\Ll}_N}

\def\PROD{\Pi}
\def\MAXW{{\mathrm{M}}}

\def\FDXDV{\, f(x,v)\dv\dx}

\def\ZNN{Z_{N,N\beta}}

\def\MAGPOT{\bar u}
\def\MAG{\bar m}
\def\HALF{\frac{1}{2}}
\def\Sph{{S}}
\def\AUXE{{{\mathcal{E}}_\beta}}
\def\ENGF{{F_\beta}}
\def\ENGFT{{\tilde F_\beta}}
\def\FLANDAU{F_{{\mathrm{LL}}}}
\begin{document}
%\draft command makes pacs numbers print
\draft
%
% Title
\title{Statistical equilibrium measures in micromagnetics}
%
% Authors and addresses
\author{Markos A. Katsoulakis}
\address{Department of Mathematics and Statistics, University of Massachusetts,
Amherst, MA 01003--3110, USA}
 \author{Petr Plech\'a\v{c}}
\address{Mathematics Institute, University of Warwick, Coventry, CV4 7AL, UK}
\date{\today}
\maketitle
\begin{abstract}
We derive an equilibrium statistical theory for the
macroscopic description of a ferromagnetic material
at  positive finite temperatures. Our formulation describes the
most-probable equilibrium macrostates that yield a coherent deterministic
large-scale picture varying at the size of the domain, as well as it
captures the effect of random spin fluctuations caused by the thermal
noise. We discuss connections of the proposed formulation
to the  Landau-Lifschitz theory and to the  studies
of domain formation based on Monte Carlo lattice simulations.
\end{abstract}
% body of paper here
\narrowtext
\section{Introduction and summary of results} \label{sec:level1}

Based on a lattice model recently used
for micromagnetic simulations at positive finite temperatures (see
\cite{Chui:PRL}) we derive a statistical equilibrium theory that describes
large-scale coherent structures (magnetic domains, domain walls)
in the presence of thermal fluctuations.
The resulting description of equilibrium states depends only on macroscopic
quantities: magnetization $m$, induced magnetic field $h$,
external magnetic field $h_e$ as well as structural properties of the
ferromagnetic material (crystalline anisotropy).
{Furthermore, our formulation captures thermally induced   
spatial, as well as spin random  fluctuations  on the lattice and
their effect in the macroscopic model.}
The Landau-Lifschitz theory (see \cite{LandauLifschitz,Brown})
has been successfully applied to modeling of magnetic microstructures in
ferromagnetic materials with crystalline anisotropy. However, the assumptions
under which the theory is derived rule out systematic description of thermal
noise and the theory is sometimes viewed as
a zero-temperature approximation. Recent advances in applications
of ultra-thin
magnetic films pointed out the importance of thermal fluctuations for studying
such effects as spin reversal, nucleation, metastability or hysteresis
(see, e.g.,\cite{Hadji}, \cite{Chui:PRB}).

{The approach proposed in this communication provides systematic treatment
of the micromagnetic lattice model at a (positive) {\it finite temperature}}
It allows for the derivation of, 
(i)  the probability distribution \VIZ{Maxwellian}
 of spins at all points of the magnetic specimen
 by incorporating the entropy, derived from the microscopic model, 
 into  the  continuum functional \VIZ{EnergyFunctional};  
(ii) a {\it macroscopic} Ginzburg-Landau type 
     free energy \VIZ{FreeEnergy}  for the average magnetization; 
(iii) the description of
      random fluctuations in configuration space 
      given by \VIZ{Maxwellian} and the derived  large deviation 
      principle \VIZ{Gibbs-ldp} respectively.
In the case of a large specimen where one observes well-developed
magnetic microstructure with many magnetic domains, our formulation
also provides an algorithm to compute spatially averaged magnetization
fields. In a follow-up publication we study stochastic
dynamics and domain formation and evolution
in the context of the statistical framework developed here.

The physics on the regular lattice $\LATT\subset\R^d$ ($d=2,3$)
with $N$ sites is defined in the standard way
by the means
of an interaction potential between spins $\sigma(x)$,
$\sigma(x')$ with values on the unit sphere $\Sph^2$
at two different sites $x$, $x'\in\LATT$.
Since we do not pursue an ab initio derivation, we assume that
the interaction potential
$U(x-x',\sigma(x),\sigma(x'))$ consists of
the following terms: $U = U_e+U_a+U_d+U_h$ with particular contributions
reflecting different types of interactions between magnetic moments.
The underlying crystallographic lattice structure is incorporated
through the interaction energy density $\Psi$ rather than by using
different lattice geometries.
The exchange energy is given by
\begin{equation}
U_e=-\frac{A}{2}\sum_{x,x'}\sum_{i,j} J((x-x')/N^{1/d}) \sigma^i(x)\sigma^j(x)\COMMA
\end{equation}
with the local mean-field interaction described by the positive function $J$.
This term
is often approximated by the nearest-neighbor interaction only. The anisotropy
energy related to the crystalline structure of the material is
defined by the energy density $\Psi$,
\begin{equation}
U_a = K \sum_{x} \Psi(\sigma(x))\PERIOD
\end{equation}
The non-local, long-range interaction between different magnetic moments
(spins) is described by the dipole-dipole interaction
\begin{equation}
U_d = \frac{g}{2}\sum_{x,x'} \sum_{i,j}  \nabla_{x_i}\nabla_{x_j}
\left(\frac{1}{|x-x'|} \right) \sigma^i(x)\sigma^j(x')\PERIOD
\end{equation}
With a slight abuse of notation we write
$\sigma(x)\sigma(x')\equiv\sum_{i,j}\sigma^i(x)\sigma^j(x')$ for
the scalar product of two unit vectors at sites $x$ and $x'$ and
points on $\Sph^2$ are represented by unit vectors with the components
$\sigma^i$, $i=1,2,3$.
We define the interaction Hamiltonian of
the system
\begin{equation}\label{Hamiltonian}
H_N(\sigma) = -\frac{1}{N^2}\sum_{x,x'\in\LATT} U(x-x',\sigma(x),\sigma(x'))\,
\PERIOD
\end{equation}
The particular scaling guarantees that for the interaction potentials
considered in the sequel the Hamiltonian remains
finite as $N \to \infty$. To simplify notation we absorb the physical
interaction parameters $A$, $g$, $K$ into definitions of corresponding
interaction potentials.
In the derivation of the statistical model
we omit the external field to keep the notation simple and focused
to the averaging procedure only. However, in the presence of an external
magnetic field $h_e$ the Hamiltonian also involves the interaction energy
\begin{equation}
U_h=-\frac{1}{N}\sum_{x}\sum_{i} h_e^i(x)\sigma^i(x)
\PERIOD
\end{equation}

The central object describing the statistical ensemble on the lattice
is the {\it canonical} Gibbs measure at the inverse temperature $\beta = \frac{1}{kT}$,
defined on the configuration space
$\Sigma=\{\sigma\SEP \sigma(x) \in \Sph^2, x \in \LATT\}$,

\begin{equation}\label{Gibbs}
P_{N,\beta}(\d\sigma) = \frac{1}{Z_{N,\beta}}
              \EXP{-\beta H_N(\sigma)}\,\PROD_N(\d\sigma)\COMMA
\end{equation}
where
\begin{equation}
Z_{N,\beta} = \int_\Sigma \EXP{-\beta H_N(\sigma)} \PROD_N(\d\sigma)
\end{equation}
denotes the partition function.
The {\it prior distribution} $\PROD_N(\d\sigma)$ models the small scale fluctuations of
the spins; we assume it is a product measure on the lattice $\LATT$
of uniform distributions on $\Sph^2$, i.e.,
that spins
at different lattice sites are independent, uniformly distributed
random variables  on $\Sph^2$.
Note that we can alternatively incorporate the anisotropy energy in
the prior distribution.
Here we study the most probable configuration on $\LATT$ according to the Gibbs
measure $P_{N,\beta}$ as $N\to\infty$. In this limit, we assume
the lattice $\LATT$ approximates a
physical domain  $\Omega\subset\R^d$.
We show
that {the  energetically most favorable  configuration
describing the large-scale features  on  $\LATT\approx\Omega$ is given by the Maxwellian distribution,} 
\begin{equation}\label{Maxwellian}
\MAXW(x,v) = \frac{1}{Z_\beta(x)} \EXP{-\beta[\Psi(v) + v (\GRAD\MAGPOT(x)
              -J*\MAG(x)-h_e)]}\COMMA
\end{equation}
{yielding the probability of having a spin $v \in \Sph^2$ at the location $x \in \Omega$.}
Here $Z_\beta(x)$ denotes the partition function. 
{The  corresponding average magnetization is
$\MAG(x) = \int_{\Sph^2} v \,\MAXW(x,v) \ddv$ with
magnetization potential $\MAGPOT$ given by $\Delta \MAGPOT = \DIV(\chi_\Omega \MAG)$.}
Furthermore
$(\MAGPOT,\MAG)$ minimizes a newly derived {\it free energy} of  Ginzburg-Landau type,
at a finite temperature:
\begin{eqnarray}\label{FreeEnergy}
\ENGF[m] =&& \int_{\R^d} \frac{1}{2}|\GRAD u|^2\ddx - \int_\Omega \HALF
(J*m)\,m\ddx + \nonumber \\
&& + \int_\Omega a^*_\beta(m)\ddx - \int_\Omega h_e m\ddx\COMMA
\end{eqnarray}
 subject to the constraint
\begin{equation}\label{Poisson}
\Delta u = \DIV(\chi_\Omega m) \,,\;\;\mbox{in $\R^d$} \PERIOD
\end{equation}
The equation \VIZ{Poisson}, involving the characteristic function
$\chi_\Omega$
of the  domain $\Omega$, is understood in the weak sense (see \cite{JamesKinderlehrer}).
%, for
%more details about the formulation see \cite{JamesKinderlehrer}.
Furthermore $J*m(x)=\int_\Omega J(x-y)m(y)dy$ and
 $$
 a^*_\beta(m) = \sup_{p\in\R^d}\{m p - a_\beta(p)\}
 $$
  is
the Legendre-Fenchel transform of the function
\begin{equation}
a_\beta(p) = \frac{1}{\beta}\log \int_{\Sph^2} \EXP{-\beta(\Psi(v) + vp)}\ddv
\PERIOD
\end{equation}

As in the case of the Boltzmann theory of dilute gases the equilibrium measure
is defined by the macroscopic quantities: the magnetic potential $\MAGPOT$
which defines the induced magnetic field $h(x)=-\GRAD\MAGPOT(x)$ and the
magnetization $\MAG(x) = \int_{\Sph^2} v \,\MAXW(x,v) \ddv$.
A related approach is also adopted in the statistical description
of coherent structures in 2D turbulence (see \cite{Turkington}).

The equilibrium measure with the density $\MAXW(x,v)$ is identified
as the minimizer of the energy functional
\begin{eqnarray}\label{EnergyFunctional}
E_\beta[f] =&& \int_{\R^d} \HALF |\GRAD u|^2 \ddx - \int_\Omega \HALF
(J*m)\,m \ddx + \nonumber \\
&& + \int_\Omega\int_{\Sph^2} \Psi(v) \,f(x,v)\ddv\ddx + \nonumber \\
&& + \frac{1}{\beta} \int_\Omega\int_{\Sph^2} f(x,v) \log f(x,v)\ddv\ddx
\PERIOD
\end{eqnarray}
Since the minimization is over the space of probability densities $f$
the following constraints must be satisfied
\begin{eqnarray}\label{c1-c3}
&& f:\Omega\times \Sph^2 \to \R^+\COMMA\;
   \int_\Omega\int_{\Sph^2} \FDXDV = 1 \COMMA \nonumber \\
&& \int_{\Sph^2} f(x,v) \dv = \frac{1}{|\Omega|} \COMMA
\end{eqnarray}
together with \VIZ{Poisson} that relates $u$ and $m$, where
\begin{equation}\label{c4}
m(x)=\int_{\Sph^2} v \,f(x,v)\dv\PERIOD
\end{equation}
The last constraint in \VIZ{c1-c3} is due to the fact that at every lattice site
there is one and only one spin.
The quantity \VIZ{c4}
evaluated at the minimizing
density $f=\MAXW$, represents the average (macroscopic) magnetization.
The last term of the energy functional is interpreted
as the thermodynamic entropy and mathematically it 
is the {\it relative entropy}  with respect to the prior distribution
of the Gibbs measure \VIZ{Gibbs}.

{We conclude the discussion of our main results with two brief comments on the random,
thermally induced fluctuations. First, \VIZ{Maxwellian} is  the probability density 
of the spin $v$ at each spatial location $x$, and upon averaging 
over the spin space  yields
$\MAG$. 
Furthermore, the large deviation principle derived below  with the rate
$E_\beta[f]$, describes the equilibrium random
fluctuations around the most probable macro-state  \VIZ{Maxwellian}
{as a function of the specimen size $N$}.
In the following sections we  outline the main steps leading
 from the microscopic Hamiltonian to the macroscopic variational
principle involving energies $E_\beta$ and $\ENGF$.
The detailed mathematical derivation is described in \cite{makpp:JSP}.

\section{Thermodynamic limit and large deviation principle}

We employ the theory of large deviations (see \cite{DupuisEllis}) to
obtain the functional $E_\beta$ from the spin Hamiltonian
\VIZ{Hamiltonian}.
We define the empirical measure on $\Nn:=\Omega\times\Sph^2$
corresponding to the spin
configurations $\sigma$,
\begin{equation}
\mu^N(\d y,\dv) = \frac{1}{N} \sum_{x\in\LATT}\delta_x(\d y)
   \delta_{\sigma(x)}(\dv)\PERIOD
 \end{equation}
{Integrating over the domain $A\times \{v\}\subset \Nn$, we readily see that
$\mu^N(A\times \{v\})$ is a coarse-grained random variable counting the number of 
spins $v$ contained in the region $A$. In the $N \to \infty$ limit it will give rise
to the probability distribution of spins in $A$.}{The empirical measure allows us  to show  that,
(i) the corresponding Gibbs states yield a coherent deterministic large scale picture,
referred to as a ``macrostate'' and varying at the size of the domain
$\Omega \times \Sph^2$, and (ii) the microscopic random  fluctuations
at the lattice cell size around the macrostates can be described explicitly
in terms of $E_\beta$.

A crucial step of the derivation
is to compute a suitable limit of the lattice Hamiltonian and
of the Gibbs measure \VIZ{Gibbs} as $N\to\infty$.
Using the empirical measure we can write the lattice Hamiltonian as
$$
\tilde H(\mu^N) = -\int_{\Nn}\int_{\Nn} U(x-x',v,v')\,
                  \mu^N(\dx,\dv)\mu^N(\dx',\dv').
$$
The partition function corresponding to the re-scaled Gibbs measure
$P_{N,N\beta}$ is
\begin{equation}
\ZNN = \int_\Sigma \EXP{-N\beta H_N(\sigma)} \PROD_N(\d\sigma)\COMMA
\end{equation}
and using
the Hamiltonian $\tilde H$ we write
$$
\frac{1}{N}\log \ZNN =
\frac{1}{N}\log \int_\Pp \EXP{-N\beta \tilde H(\mu)} \, \PROD_N(\mu^N=\mu)\d\mu
\COMMA
$$
where $\Pp$ is the set of all probability
measures $\mu$ on $\Nn$ that satisfy \VIZ{c1-c3}
(observe that if $\mu^N$ converge in the sense of measures as
$N\to\infty$ to a probability measure on $\Nn$ with  density $f$
then $f$ satisfies the constraints \VIZ{c1-c3}).
% Alternatively we can write
% $$
% \frac{1}{N}\log \ZNN =
% \frac{1}{N}\log \int_\Pp \EXP{-N\beta \tilde H(\mu)} \, \PROD_N(\d\mu)
% \PERIOD
% $$
Using Sanov's Theorem  and
an auxiliary coarse grained process (see \cite{BET})
we can show the large deviation principle for the prior distribution
with the {\it rate functional} $S(\mu)$:
\begin{equation}
\PROD_N(\mu^N=\mu) \approx \EXP{-N S(\mu)}\COMMA
 \end{equation}
  as $N\to\infty$.
Here
$$
S(\mu) = \int f(x, v) \log f(x, v) \,\dx\dv\,  \quad \mbox{if \,$\d\mu=f\dx\dv$}
\COMMA
$$
and $S(\mu)=\infty$ otherwise, which  is known in
probability theory as the  {\it relative entropy} with respect to the uniform measure
$\dx\dv$
\cite{DupuisEllis}.
For the rigorous meaning of the approximation sign above
in the framework of the theory of large deviations,  we refer to \cite{DupuisEllis}.
By employing  the asymptotics  of $\PROD_N(\mu^N=\mu)$,
 as well as the Laplace principle, \cite{DupuisEllis},
we compute the limit as $N\to\infty$
\begin{eqnarray}\label{eq14}
\lim_{N \to \infty}\frac{1}{N}\log \ZNN 
 & = & \lim_{N \to \infty} \frac{1}{N} \log\int_{\Pp} 
       \EXP{-N(\beta \tilde H(\mu)+S(\mu))} \d\mu \nonumber \\
 & = &  -\inf_{\mu\in\Pp}\{\beta \tilde H(\mu)+S(\mu)\}
\PERIOD
\end{eqnarray}
Similarly we obtain the large deviation principle for the Gibbs measure, as $N\to\infty$,
\begin{equation}\label{Gibbs-ldp}
P_{N,N\beta}(\mu^N=\mu) \approx \EXP{-N[\beta\tilde H(\mu) + S(\mu)-\inf_{\mu}\{\beta\tilde H(\mu) + S(\mu)\}]}
%_{N,N\beta}(\mu^N \in \mu) \approx \EXP{-N[\beta\tilde H(\mu) + S(\mu)-
%                    \inf_{\mu}\{\beta\tilde H(\mu) + S(\mu)\}]}\d\mu
\PERIOD
\end{equation}
The rigorous derivation is technically involved due to the
nature of the singularity in the dipole-dipole interaction (see also
\cite{JamesMuller}). We address rigorously this issue in \cite{makpp:JSP}.

The expression \VIZ{Gibbs-ldp} is interpreted as follows:
the most probable configuration $\mu$ of the Gibbs measure is the
{\it minimizer} of $\beta\tilde H(\mu) + S(\mu)$, yielding
the  large scale structure (at the size of the domain
$\Nn$) at equilibrium. As we show below, this minimizer turns out to
be the probability density \VIZ{Maxwellian}.
Note here that $\beta\tilde H(\mu) + S(\mu)$ is finite if and only if
the measure $\mu$ has a density $f$,  in which case
we define the energy functional
$$
E_\beta[f]= \tilde H(\mu) + \frac{1}{\beta} S(\mu)\, .
$$
 Substituting for
expressions on the right hand side and using \VIZ{Poisson}
we immediately obtain \VIZ{EnergyFunctional}.

Finally,  \VIZ{Gibbs-ldp} captures the
microscopic {\it spatial} random fluctuations of the empirical measure $\mu^N$
as a function of $N$ (i.e., the total number  of spins in the specimen),
around the most probable macrostate given by the minimizer \VIZ{Maxwellian}
of \VIZ{EnergyFunctional}.

\section{Variational principle}
We show that the Maxwellian  $\MAXW(x,v)$ which represents the equilibrium
macrostate of the microscopic system is the minimizer of the energy
functional $E_\beta$. Using  \VIZ{Poisson} we define an
auxiliary functional,
\begin{eqnarray}\label{AuxEnergy}
\AUXE[f, u]=&& -\int_{\R^d} \frac{1}{2}|\GRAD u|^2\dx -
                 \int_\Omega \HALF (J*m)\,m\ddx + \nonumber \\
&& + \int_\Omega\int_{\Sph^2} (\Psi(v)+v \GRAD u) \FDXDV + \nonumber \\
&& + \frac{1}{\beta}\int_\Omega\int_{\Sph^2} f(x,v)\log f(x,v) \dx\dv \PERIOD
\end{eqnarray}
We observe that $E_\beta[f] = \sup_{u\in H^1(\R^d)} \AUXE[f,u]$, and
consequently
$$
\inf_{f}E_\beta[f] = \inf_{f}\sup_{u} \AUXE[f,u] \geq
\sup_u\inf_f \AUXE[f,u]\PERIOD
$$
Note that the infimum is taken over the densities
$f$ that satisfy constraints \VIZ{c1-c3} and also relations
\VIZ{Poisson} together with \VIZ{c4}. For a fixed $u$ we can
construct the minimizer of $\AUXE[f,u]$ over $f$ by varying the probability
density $f$ using suitable push-forward maps (see
\cite{Jordan,JordanKinderlehrerOtto}) that
preserve the constraints \VIZ{c1-c3}. Then the minimizer $\MAXW_u$,
$\AUXE[\MAXW_u,u] = \min_f \AUXE[f,u]$ for given, fixed $u$ satisfies
\begin{eqnarray}\label{MaxwellianU}
\MAXW_{u}(x,v) =&& \frac{1}{Z_\beta(x)}
      \EXP{-\beta[\Psi(v) + v (\GRAD u(x)- J*m_u(x))]}\COMMA
\end{eqnarray}
where $Z_\beta(x)$ is the corresponding partition function and
$m_u(x) = \int_{\Sph^2} v\,\MAXW_u(x,v) \ddv$.
Note that $a_\beta(\GRAD u - J*m_u) = 1/\beta \log Z_\beta(x)$.
Straightforward calculations show that
\begin{equation}\label{Mueq}
m_u(x) = \DIFF_p a_\beta ( J*m_u-\GRAD u) \PERIOD
\end{equation}
The existence of $m_u$ satisfying \VIZ{Mueq} follows from minimizing the
functional
$$
 -\int_\Omega \HALF (J*m)\,m\ddx + \int_\Omega m \GRAD u \ddx
+ \int_\Omega a^*_\beta(m)\ddx\, .
$$
Hence we obtain that
\begin{eqnarray*}
&& \AUXE[\MAXW_u,u] =  \min_f\AUXE[f,u]=
-\int_{\R^d}\HALF|\GRAD u|^2\ddx + \\
 + && \int_\Omega \HALF (J*m_u)\,m_u\ddx
      -\int_\Omega a_\beta(\GRAD u - J*m_u)\ddx
\PERIOD
\end{eqnarray*}
Using the duality between $a_\beta$ and $a^*_\beta$ we have
$$
\sup_u\inf_f \AUXE[f,u] = \sup_u\AUXE[\MAXW_u,u] =
\inf_{m\,\&\,\VIZ{Poisson}} \ENGF[m]\PERIOD
$$
Closer inspection of the functional $\ENGF$ proves that the last infimum
is attained and that a minimizer $\MAG$ satisfies
$$
\MAG = \DIFF_p a_\beta(J*\MAG - \GRAD\MAGPOT)
$$
and $\MAG$, $\MAGPOT$
are related by \VIZ{Poisson}. The expression \VIZ{MaxwellianU} also implies
the equilibrium Maxwellian $\MAXW(x,v)$ given by \VIZ{Maxwellian}. Using the
constructed fields $\MAG(x)$, $\MAGPOT(x)$ and $\MAXW(x,v)$
we can check that, in fact,
\begin{eqnarray*}
\inf_f E_\beta[f] = \inf_f\sup_u \AUXE[f,u] = && \sup_u\inf_f\AUXE[f,u] =\\
 = && \min_{m\,\&\,\VIZ{Poisson}}\ENGF[m]  = \ENGF[\MAG]\PERIOD
\end{eqnarray*}
Thus the minimizer of the new
Ginzburg-Landau energy $\ENGF[m]$ provides the minimum value of the energy
functional $E_\beta[f]$ as well as it describes, through the Maxwellian
density $\MAXW(x,v)$ the structure of the most-probable macrostate.
The rigorous treatment of all steps outlined in the previous sections is described 
in \cite{makpp:JSP}.

\section{Discussion}

A natural question is how the presented model relates to the Landau-Lifschitz
theory (\cite{LandauLifschitz,Brown}) and how it can be interpreted in view
of recent studies of domain formation based on Monte Carlo simulations
(\cite{Chui:PRL,Chui:PRB}). First we address the former issue: we expand
the convolution $J*m$ in the free energy \VIZ{FreeEnergy}, $J*m = J_0 m +
J_2/2\, \LAP m + \dots$, where $J_0 = \int_\Omega J(r)\dd r$,
$J_2=\int_\Omega |r|^2 J(r) \dd r$, and after substituting into
\VIZ{FreeEnergy} we obtain
\begin{eqnarray}\label{FreeEnergyTF}
\ENGFT[m] = && \int_\Omega \frac{J_2}{2} |\GRAD m|^2 \ddx +
                 \int_{\R^d}\HALF |\GRAD u|^2 \ddx + \nonumber \\
              && \int_\Omega ( a^*_\beta(m) - \frac{J_0}{2}|m|^2 )\ddx
                 - \int_\Omega h_e m \ddx \COMMA
\end{eqnarray}
which can be interpreted as the finite-temperature analogue of the
Landau-Lifschitz free energy:
\begin{eqnarray}\label{FreeEnergyLL}
F_{{\mathrm{LL}}}[m] = && \int_\Omega \frac{A}{2} |\GRAD m|^2 \ddx + 
                 \int_{\R^d}\HALF |\GRAD u|^2 \ddx +  \nonumber \\
              && +\int_\Omega \Psi(m) \ddx - \int_\Omega h_e m \ddx \PERIOD
\end{eqnarray}
However, the free energy functional $\FLANDAU$ is minimized subject to the
non-convex constraint $|m|=1$. In the formulation presented here the
direct calculation implies that since $|\DIFF_p a_\beta(p)|\leq 1$
we have $a^*_\beta(m)=\infty$ for $|m|>1$ and consequently the energy
$\ENGFT$ (or $F_\beta$) is minimized subject to the constraint $|m|\leq 1$. This remarkable
difference from the Landau-Lifschitz theory is caused
by the presence of the thermal
fluctuations and the averaged nature of the magnetization $m$ in our
model. We conclude the discussion on the relation
of our proposed finite temperature model to
the Landau-Lifschitz theory with an interesting point: one would
be tempted to say that a suitable zero-temperature limit  ($\Gamma$-limit) of the
free energy $\ENGF$
%($\Gamma$-limit
%of the functionals $\ENGF$)
should yield the Landau-Lifschitz model
or its relaxation.
Indeed, the $\beta \to \infty$ limit  of the
free energy $\ENGF$ or $\ENGFT$ can be explicitly calculated, at least formally,
yielding   in \VIZ{FreeEnergy} and \VIZ{FreeEnergyTF}:
$$
a_\infty^*(m) =
\sup\big\{f:\R^n \mapsto \R, f \le \Psi\, \mbox{on}\,  \Sph^2\, , f\, 
\mbox{convex}\big\}\COMMA
$$
if  $|m| \le 1$ and $a_\infty^*(m)=\infty$, if $|m|>1$.
We note that the same
energy relaxation was obtained in \cite{DeSimone}
by a direct minimization of   \VIZ{FreeEnergyLL} over all admissible Young
measures, when the energy exchange term $A/2 |\GRAD m|^2$ is neglected.

Existence of magnetic domains and different types of magnetic walls
in the Landau-Lifschitz theory is attributed to the competition of
different contributions in the free energy and to the non-convex
constraint $|m|=1$. Due to the thermal agitation incorporated in the
finite-temperature free energy $\ENGF$ (or $\ENGFT$) the norm $|m|$
also fluctuates. Although for low temperatures $|m|$ can be close to
the unit sphere, in general, we have the convex constraint $|m|\leq 1$.
Nevertheless, our model allows for domain formation at finite-temperatures
whenever the exchange energy is strong enough compared to the
temperature.

More specifically, in the absence of an external field $h_e$, the condition
that $I - J_0 \DIFF^2_{pp} a_\beta(0)$ is positive definite guarantees
that $m\equiv 0$ and $u\equiv 0$ is the minimizer of  $\ENGF$ (as well as $\ENGFT$)
implying a uniform
state, i.e., no domain formation. When the above condition is violated,
then for a suitable domain $\Omega$ and anisotropy,
there exist non-trivial solutions
$m \ne 0$ and $u\equiv 0$ to the
Lagrange-Euler equation for the energy $\ENGFT$ (or $\ENGF$),
\begin{equation}\label{LEeqFT}
J_2\LAP m = - J_0 m + \DIFF_p a^*_\beta(m)\COMMA
\end{equation}
which have lower energy than the uniform state $m\equiv 0$, $u\equiv 0$.
Similar steady states on 2D lattices, referred to as many-soliton
solutions, were
first analytically predicted and also observed in Monte Carlo simulations
in \cite{Chui:PRL}.
Under simplifying assumptions on the geometry of the domain it is
not difficult to calculate the domain-wall profile, as well as its dependence
on temperature, the exchange energy and anisotropy energy directly
from \VIZ{LEeqFT}.
Note that since the magnetization $m$ in our model is computed
by averaging over the thermal fluctuations of spins the domains are identified
with regions in $\Omega$ where $|m(x)|\approx |m^k_\beta|\leq 1$ and
$m^k_\beta$ are non-trivial constant solutions of \VIZ{LEeqFT}.
In a forthcoming publication,
domain formation and evolution, as well as the
corresponding stochastic dynamics
are studied both numerically and analytically  in the context
of the  statistical theory developed here.

\medskip

\noindent
{\bf Acknowledgments:}
The research of M.A.K. is partially supported by the National Science Foundation
through DMS-0100872 and  DMS-9626804. The visits of P.P. at the University of Massachusetts
were partially funded by  DMS-9626804 and DMS-0100872.

\bibliographystyle{prsty}
\bibliography{makpp}

%\begin{references}
%\bibitem{tag} Fake bibitem.
%\end{references}

% \begin{figure}
% \caption{}
% \label{}
% \end{figure}

% \begin{table}
% \caption{}
% \label{}
% \end{tabular}
% \end{table}

\end{document}